\documentclass[prb,twocolumn,superscriptaddress,amsmath,amssymb,aps,longbibliography]{revtex4-2} 

\usepackage[T1]{fontenc}
\usepackage[latin9]{inputenc}
\usepackage{amssymb,amsfonts,amsmath,amsthm,mathrsfs,bbm,bm} 
\usepackage{graphicx}
\usepackage{color}
\usepackage{float}
\usepackage{indentfirst}
\usepackage{subfigure}
\usepackage{multirow}
\usepackage{tabu}
\usepackage{threeparttable}
\usepackage{booktabs}
\usepackage{txfonts}
\usepackage[normalem]{ulem}
\usepackage[euler]{textgreek}

\usepackage[dvipsnames]{xcolor}
\usepackage{soul}
\usepackage[colorlinks=true,urlcolor=MidnightBlue,citecolor=MidnightBlue,linkcolor=MidnightBlue]{hyperref}

\usepackage{adjustbox}
\usepackage[capitalise]{cleveref}

\newcommand{\vx}{{\mathbf{x}}}

\crefname{algocf}{alg.}{algs.}
\Crefname{algocf}{Algorithm}{Algorithms}

\begin{document}
\title{Accelerating two-dimensional tensor network optimization by preconditioning}

\author{Xing-Yu Zhang}
\email{Xingyu.Zhang@Ugent.be}
\thanks{These authors contributed equally to this work.}
\affiliation{Department of Physics and Astronomy, Ghent University, Krijgslaan 281, 9000 Gent, Belgium}

\author{Qi Yang}
\email{qiyang@mail.ustc.edu.cn}
\thanks{These authors contributed equally to this work.}
\affiliation{Institute for Theoretical Physics Amsterdam and Delta Institute for Theoretical Physics, University of Amsterdam, Science Park 904, 1098 XH Amsterdam, The Netherlands}

\author{Philippe Corboz}
\affiliation{Institute for Theoretical Physics Amsterdam and Delta Institute for Theoretical Physics, University of Amsterdam, Science Park 904, 1098 XH Amsterdam, The Netherlands}

\author{Jutho Haegeman}
\affiliation{Department of Physics and Astronomy, Ghent University, Krijgslaan 281, 9000 Gent, Belgium}

\author{Wei Tang}
\email{wei.tang.phys@gmail.com}
\affiliation{Department of Physics and Astronomy, Ghent University, Krijgslaan 281, 9000 Gent, Belgium}

\begin{abstract}
    We revisit gradient-based optimization for infinite projected entangled pair states (iPEPS), a tensor network ansatz for simulating many-body quantum systems. This approach is hindered by two major challenges: the high computational cost of evaluating energies and gradients, and an ill-conditioned optimization landscape that slows convergence. To reduce the number of optimization steps, we introduce an efficient preconditioner derived from the leading term of the metric tensor. We benchmark our method against standard optimization techniques on the Heisenberg and Kitaev models, demonstrating substantial improvements in overall computational efficiency. Our approach is broadly applicable across various contraction schemes, unit cell sizes, and Hamiltonians, highlighting the potential of preconditioned optimization to advance tensor network algorithms for strongly correlated systems.
\end{abstract}

\maketitle

\section{Introduction}
Accurately describing and understanding strongly correlated quantum matter remains a central challenge in condensed matter physics. Tensor network states, such as matrix product states (MPS) and projected entangled-pair states (PEPS) have emerged as powerful frameworks for numerically simulating strongly correlated quantum lattice systems, due to their ability to efficiently capture the relevant entanglement structures~\cite{cirac-matrix-2021,xiang-density-book-2023}. In particular, infinite PEPS (iPEPS) provide a natural framework for studying two-dimensional lattice models directly in the thermodynamic limit~\cite{verstraete2004renormalization}.

As tensor networks constitute nonlinear classes of variational trial states, accurate ground state aproximations are obtained by applying numerical optimization techniques. The methodology has evolved from early imaginary-time evolution techniques~\cite{PhysRevLett.101.090603, PhysRevB.81.165104, PhysRevB.92.035142} to the more recent application of conventional gradient-based optimization algorithms~\cite{corboz2016variational,vanderstraeten2016gradient, liao2019differentiable, 10.21468/SciPostPhys.10.1.012, PhysRevB.108.085103}. 
These methods minimize the energy functional directly, leading to lower variational energies and more accurate states. 
Furthermore, the procedure for computing the necessary energy gradients is greatly simplified by automatic differentiation (AD)~\cite{liao2019differentiable}, which makes the application of gradient-based optimization algorithms for iPEPS much more feasible and flexible~\cite{liao2019differentiable, 10.21468/SciPostPhys.10.1.012, PhysRevB.108.085103, 10.21468/SciPostPhysLectNotes.86, richards2025,PhysRevResearch.7.013237, 10.21468/SciPostPhysCodeb.52, brehmer_2025_16755150}.

Nonetheless, the gradient-based optimization of iPEPS still faces many technical challenges in practice.
In particular, they are generally considered to be much more computationally expensive than imaginary-time evolution methods, which has limited their broader application. 
There are two main reasons for this.
First, variational optimization requires computing the energy and its derivative at each iteration step.
This requires one to perform an infinite-tensor network contraction at every step, which is very expensive~\cite{nishino1997corner, PhysRevB.80.094403, zauner2018variational,fishman2018faster, 10.21468/SciPostPhysLectNotes.7}.
The significant memory and computational costs associated with these contractions largely limit the achievable virtual bond dimension in iPEPS simulations. 
Second, the energy landscape in the iPEPS optimization is often ill-conditioned, leading to slow convergence of the optimization process and sensitivity to initial conditions. 
In practice, the number of iterations required for convergence grows rapidly with the number of parameters, which becomes formidable when a large bond dimension is used.
More specifically, in quasi-Newton algorithms such as L-BFGS, which are routinely employed in gradient-based optimizations of iPEPS, the optimizer aims to construct an approximation of the (inverse) Hessian to accelerate convergence. However, the ill-conditioned optimization landscape hinders the acquisition of an accurate approximation to the inverse Hessian, resulting in a significantly slower optimization process.
Together, these two factors make the gradient optimization of iPEPS very expensive, highlighting the need for methods that can effectively reduce the number of iterations in the optimization.

Among others, preconditioning is a widely employed technique to improve convergence. 
Preconditioning can be interpreted as a transformation of the parameter space, which, when properly chosen, mitigates the ill-conditioning of the optimization landscape and thereby substantially accelerates convergence of the optimization. 
However, an appropriate choice of preconditioner is problem-dependent and often far from obvious.
In the context of tensor network states, Ref.~\cite{10.21468/SciPostPhys.10.2.040} proposes that a natural choice for the preconditioner is the metric of the tangent space of the corresponding tensor network state manifold, as it effectively encodes the geometric information of the tensor network state.  
Such preconditioner has been shown effective in the optimization of infinite MPS~\cite{10.21468/SciPostPhys.10.2.040}, multi-scale entanglement renormalization ansatz~\cite{10.21468/SciPostPhys.10.2.040}, and continuous MPS~\cite{PhysRevB.108.035153}.
It is worth mentioning that this choice of preconditioner is conceptually connected to a variety of existing numerical methods, such as the time-dependent variational principle in quantum systems~\cite{dirac1930note,RevModPhys.44.602,PhysRevLett.107.070601}, the Gauss-Newton method in non-linear least-squares problems~\cite{Chen2011HessianMV}, the stochastic reconfiguration in variational Monte Carlo~\cite{PhysRevLett.80.4558}, and the natural gradient in machine learning \cite{6790500, stokes2020quantum}.

In this work, we employ the preconditioning scheme from Ref.~\cite{10.21468/SciPostPhys.10.2.040} to the gradient-based optimization of iPEPS.
Since constructing the full iPEPS tangent-space metric is computationally expensive, we also investigate the use of a local approximation to the metric as preconditioner. 
We show that employing the preconditioner can substantially accelerate the energy minimization procedure, for both single- and large-unit-cell iPEPS optimizations. 
This demonstrates the power and practicality of the tangent-space-based preconditioner in improving the efficiency of iPEPS optimization.

The manuscript is organized as follows. \Cref{sec:preconditioner-basics} presents a brief introduction to preconditioning in numerical optimizations. \Cref{sec:Preconditioner in iPEPS optimization} details the application of our preconditioner to iPEPS optimization and its technical implementation. 
In Sec.~\ref{sec:Numerical results}, for the Heisenberg and Kitaev models, we compare the performance of preconditioned optimization with standard optimization, using both single- and large-unit-cell iPEPS.

\section{Basics of preconditioning in optimization} \label{sec:preconditioner-basics}
In this section, we briefly review the concept of preconditioning in the context of optimization algorithms. 
For simplicity, we consider the minimization problem of the following quadratic function
\begin{equation}
    f(\vx) = \frac{1}{2}{\vx}^T A \vx -b^T \vx
\end{equation}
where $A$ is a positive definite matrix and $b$ is a vector.
The solution to this problem is  $\vx^* = A^{-1}b$. 
The performance of standard gradient-based optimization algorithms in this problem strongly depends on the condition number of Hessian matrix~$A$. 
A large condition number of $A$ leads to an ill-conditioned optimization landscape in which it is hard to navigate, as the level curves of the objective function $f$ form a steep and narrow valley, causing slow convergence.

A key observation is that the optimization landscape changes under linear transformations of $\vx$. 
More specifically, one can consider a linear transformation represented by a invertible matrix $Q$, under which the variables are transformed as $\tilde\vx = Q \vx$. 
In terms of the new variables, the cost function becomes 
\begin{equation}
\tilde{f}(\tilde{\vx}) = \frac{1}{2} \tilde{\vx}^T [(Q^T)^{-1} A Q^{-1}] \tilde{\vx} - (b^T Q^{-1}) \tilde{\vx}.
\end{equation}
If the condition number of the new Hessian matrix $(Q^T)^{-1} A Q^{-1}$ is reduced, the optimization with the new variables $\tilde{x}$ becomes more efficient. 

In practice, there is no need to transform the variables explicitly. 
Instead, we only need to properly utilize the information from the new optimization problem. 
More specifically, in the new cost function, the search direction is determined from the gradient 
\begin{equation}
\tilde{g} = \mathrm{d}_{\tilde{\vx}} \tilde{f} = [(Q^T)^{-1} A Q^{-1}] \tilde{\vx} - (Q^T)^{-1} b.
\end{equation}
In terms of the original variables $\vx$, the search direction becomes 
\begin{equation}
Q^{-1} \tilde{g} = P^{-1} ( A \vx - b) = P^{-1} \mathrm{d}_{\mathbf{x}} f ,
\end{equation}
where $P = Q^T Q$. 
The positive definite matrix $P$ is known as the preconditioner, and $P^{-1} \mathrm{d}_{\mathbf{x}} f$ is called the preconditioned gradient. 
Remarkably, if one chooses the preconditioner as the Hessian itself $P = A$, the preconditioned gradient $g^\prime$ becomes
\begin{equation}
    g^\prime = P^{-1} g = A^{-1}(A\vx - b) = \vx - \vx^*.
    \label{eq:pregrad}
\end{equation}
In this case, even the simplest steepest descent method can converge to the optimal solution $\vx^*$ in one step.

While the analysis above assumes a quadratic cost function, the logic of preconditioning extends naturally to general nonlinear optimization problems. 
Although there is no rigorous relation between the convergence rate and the condition number of the Hessian in such cases, a large condition number  of the Hessian typically leads to an ill-conditioned optimization landscape and hinders convergence of the optimization. 
In these cases, one may consider employing preconditioning to mitigate this issue and accelerate optimization, where in this case the preconditioner $P$ can itself become dependent on the the position $\vx$~\footnote{In the nonlinear case, the position-dependent preconditioner can still be related to a coordinate transformation as $P(\vx) = Q(\vx)^T Q(\vx)$, where now $Q(\vx)$ is the Jacobian of a general nonlinear coordinate transformation.}.
Once a particular form of the preconditioner is specified, its incorporation into standard quasi-Newton algorithms is well established. In quasi-Newton methods, the algorithm maintains an approximation to the Hessian. It is then important to incorporate the preconditioning step consistently within the Hessian approximation process. 
For further details, one may refer to standard references in numerical optimization, e.g., Ref.~\cite{wright1999numerical}.

\section{Preconditioning in iPEPS optimization}
\label{sec:Preconditioner in iPEPS optimization}
\subsection{Full and local metric preconditioner}
In the context of iPEPS optimization, we aim to minimize the energy functional $E(A)$ with respect to the parameters in the local tensor $A$ of iPEPS. The energy functional is given by
\begin{equation}
    E(A) = \frac{\langle \psi(\bar{A}) | H | \psi(A) \rangle}{\langle \psi(\bar{A}) | \psi(A) \rangle},
\end{equation}
where $H$ is the Hamiltonian of the system and $| \psi(A) \rangle$ is the wave function represented by the iPEPS ansatz. To get the energy, we need to contract the tensor network, which can be done approximately using methods such as the corner transfer matrix renormalization group (CTMRG)~\cite{nishino1997corner, PhysRevB.80.094403} or variational uniform matrix product state (VUMPS)~\cite{zauner2018variational, 10.21468/SciPostPhysLectNotes.7}. Then the energy functional can be expressed as a function of the tensor $A$ and the environment tensors obtained from the contraction. Without loss of generality, we consider a system on a square lattice 
that includes only nearest-neighbor interactions, the energy of which can be computed as 
\begin{equation}
    E(A) =\adjincludegraphics[valign=c,width=0.15\textwidth]{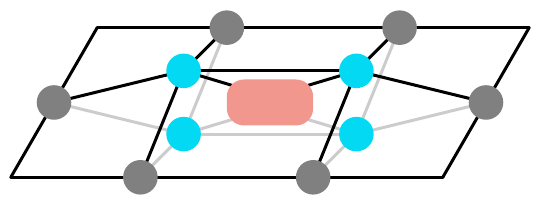}/\adjincludegraphics[valign=c,width=0.15\textwidth]{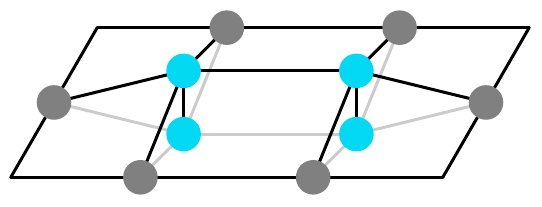},
    \label{equ:energy}
\end{equation}
where $A=\adjincludegraphics[valign=c,width=0.03\textwidth]{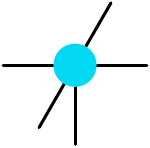}$ is the local PEPS tensor, the red rectangle represents the local Hamiltonian tensor, and the gray circles represent the environment tensors, respectively. The gradient of the energy functional with respect to the tensor $A$ can be computed using backward mode AD techniques~\cite{liao2019differentiable}. 

Once both the energy $E(A)$ and its derivative $\partial_A E(A)$ are obtained, standard gradient-based optimization methods can be employed to find the best approximation to the ground state within the iPEPS manifold.
However, as mentioned earlier, the optimization landscape is often ill-conditioned, typically requiring a large number of iterations for the optimization to reach convergence.
The ill-conditioning of the optimization landscape can largely be attributed to the geometry of the iPEPS manifold as it is embedded in the full Hilbert space. The Hilbert space inner product induces a non-trivial inner product in the tangent space at each point of the manifold, i.e.~a nontrivial metric.
This motivates the use of this tangent-space metric as a preconditioner for the optimization of the tensor $A$, thereby incorporating the geometry of the iPEPS manifold and mitigating the ill-conditioning issue~\cite{10.21468/SciPostPhys.10.2.040}.

Pure gradient descent with this preconditioner becomes directly related to imaginary-time evolution according to the time-dependent variational principle~\cite{dirac1930note,RevModPhys.44.602,PhysRevLett.107.070601}. Indeed, the preconditioned gradient can be interpreted as the tangent space projection of the full energy gradient in Hilbert space. However, as compared to full-fledged Riemannian optimization on the iPEPS manifold, which also requires suitable choices of retraction and vector transport, incorporating the metric as a preconditioner on the linear space of the tensors $A$ directly offers a number of practical advantages. In particular, merely using the metric as a preconditioner allows further modifications (described in the next subsections) that would not be possible if the metric is used as the true inner product.

In the following, we discuss the construction of this preconditioner in the context of iPEPS optimization.   
Given an iPEPS wavefunction $|\psi(A)\rangle$, the tangent vector is defined as~\cite{10.21468/SciPostPhysLectNotes.7,PhysRevB.92.201111} 
\begin{equation}
    |\psi(A; B) \rangle = \sum_i B^i \partial_{A^i} |\psi(A)\rangle =  \sum_i B^i |\partial_i \psi(A)\rangle,
    \label{eq:tangent-vector}
\end{equation}
where $A^i$ stands for the entry in tensor $A$. 
This tangent vector has been extensively used in excited state calculations of iPEPS~\cite{PhysRevB.99.165121, PhysRevB.101.195109, 10.21468/SciPostPhys.12.1.006, PhysRevB.108.195111}.
The tangent space metric $\mathbf{N}$ is defined using the physical overlap between the states represented by the tangent vectors~\cite{10.21468/SciPostPhysLectNotes.7} 
\begin{equation}
\sum_{i j} (\bar{B}^\prime)^i N_{ij} B^j = \langle \psi(\bar{A}; \bar{B}^\prime) | \psi(A; B) \rangle 
\end{equation}
Formally, the metric can be represented as $N_{i j} = \langle \partial_i \psi(\bar{A}) |  \partial_j \psi(A) \rangle$, where $|\partial_j \psi(A) \rangle$ is an infinite sum of iPEPS states with a single vacancy, summed over all possible vacancy locations.
Therefore, the metric $\mathbf{N}$ consists of infinite number of terms, which can be represented as
\begin{equation}
    \begin{aligned}
        \mathbf{N} &= \adjincludegraphics[valign=c,height=0.018\textheight]{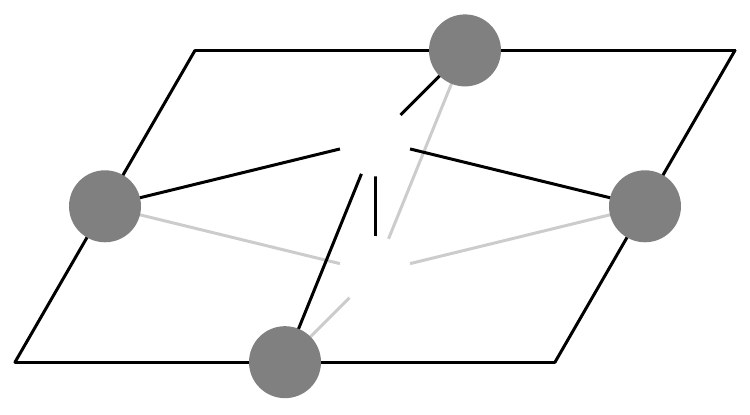} + \adjincludegraphics[valign=c,height=0.018\textheight]{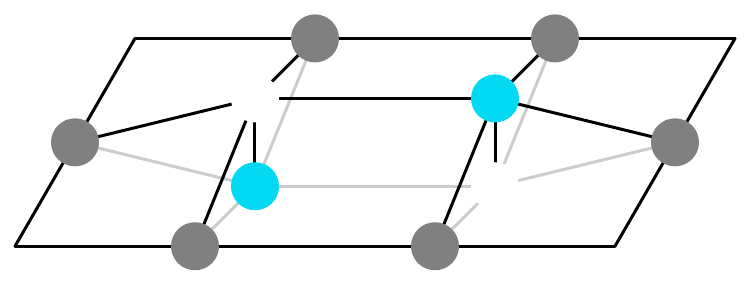} + \adjincludegraphics[valign=c,height=0.018\textheight]{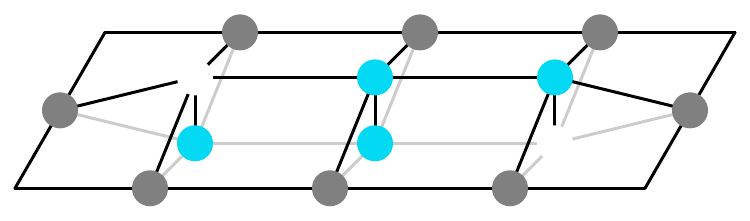} + \\
        & \adjincludegraphics[valign=c,height=0.03\textheight]{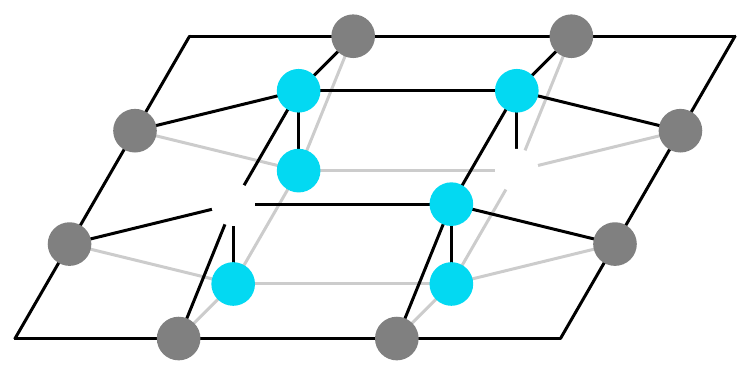} + \adjincludegraphics[valign=c,height=0.03\textheight]{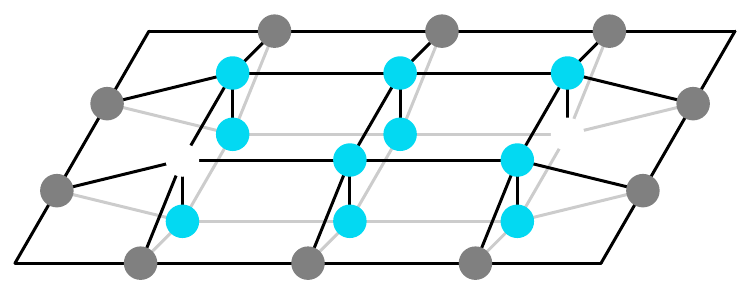} + \cdots.
    \end{aligned}
    \label{equ:full_metric}
\end{equation}
To construct the metric $\mathbf{N}$, one can compute the derivative $\partial_{\bar{A}^\prime} \langle \psi(\bar{A}^\prime) | \partial_j \psi(A) \rangle|_{A^\prime=A}$, 
where $ \langle \psi(\bar{A}^\prime) | \partial_j \psi(A) \rangle$ can be computed explicitly while its derivative with respect to $\bar{A}^\prime$ can be performed with forward mode AD.

Since the metric \eqref{equ:full_metric} consists of infinitely many terms, each of which take the form of a two-point correlation function, its computation requires a computationally expensive approximation, especially when the iPEPS has a large bond dimension.
To reduce the computational cost, we therefore consider an approximation to the full metric. 
A natural choice is to retain only the first term in Eq.~\eqref{equ:full_metric}, i.e., 
\begin{equation}
    \mathbf{N_\mathrm{local}}=\adjincludegraphics[valign=c,height=0.018\textheight]{figures/FullP11.pdf},
    \label{equ:local_metric}
\end{equation}
which we refer to as the local metric.
This approximation corresponds to the single-site environment of a iPEPS tensor, which is readily available as soon as we compute the iPEPS norm. 

A further significant reduction of the computational cost of the preconditioning step can be obtained by setting the virtual bond dimension of the environment tensors in Eq.~\eqref{equ:local_metric} to $1$. 
The environment with $\chi=1$ can also be obtained with the belief propagation (BP) algorithm~\cite{bethe1935statistical, doi:10.1142/0271, yedidia2003understanding, felzenszwalb2006efficient}, and is equivalent to a mean-field approximation of the environment. In this case, applying the inverse preconditioner becomes almost trivial.

\subsection{Regularization of the preconditioner}
In the definition of the tangent vector \eqref{eq:tangent-vector}, directions $B$ corresponding to infinitesimal gauge transformations give rise to tangent vectors $|\Psi(A; B)\rangle$ that are zero states in the Hilbert space, and thus zero eigenvectors of the full tangent-space metric. This is no longer true for the local approximation of the tangent space metric. However, we find that, even if the gauge redundancies are removed, both the full metric in Eq.~\eqref{equ:full_metric} and the local approximation in Eq.~\eqref{equ:local_metric} may still be nearly singular, especially in the early stages of optimization.

It is therefore advisable to always add a regularization term $\delta I$ to the preconditioner~\cite{10.21468/SciPostPhys.10.2.040}, where $I$ is the identity matrix and $\delta$ is a small positive constant~\footnote{In the context of the nonlinear least-squares problem, an analogous modification of the Gauss-Newton algorithm is known as the Levenberg-Marquardt algorithm~\cite{levenberg1944method}.}.
In practice, we find that it is often beneficial to use a relatively large value of $\delta$ in early stages of the optimization, and much smaller values when the iPEPS state is close to the optimal point. 
We tested two adaptive schemes: setting $\delta$ to the energy difference between steps, $|E_{i+1} - E_i|$, or to the squared gradient norm, $|g|^2$. 
In both schemes, the value of $\delta$ automatically decreases as the optimization progresses, and both approaches yield comparable results.

\subsection{Computation of the preconditioned gradient}
In practice, explicitly constructing the preconditioner and computing its inverse can become very expensive when the iPEPS bond dimension is large. 
This can be avoided by implicitly solving the linear system $P g' = g$, which only requires matrix-vector multiplications. 
This raises another problem that $P$ typically has a very large condition number, which makes the convergence of the linear system extremely slow.  
Nevertheless, we find that a rough approximation, e.g., the solution obtained from a single outer iteration of GMRES with 30 inner iterations (Krylov dimension), is often sufficient to achieve a substantial performance boost in the optimization process. 
It remains to be explored, however, how to solve the linear system more accurately without introducing significant additional computational cost.

\section{Numerical results}
\label{sec:Numerical results}
\subsection{Results in iPEPS with single unit cell}
\label{sec:localfull}
To demonstrate the effectiveness of the preconditioner introduced in Sec.~\ref{sec:Preconditioner in iPEPS optimization}, we perform numerical experiments in the two-dimensional quantum Heisenberg model on a square lattice.
More specifically, we compare the performance of using the full metric \eqref{equ:full_metric} and the local metric \eqref{equ:local_metric} as preconditioners.
The Hamiltonian of the model is given by
\begin{equation}
    H = \sum_{\langle i,j \rangle} \mathbf{S}_i \cdot \mathbf{S}_j, \label{eq:H_heisenberg}
\end{equation}
where $\mathbf{S}_i$ is the spin-$\frac{1}{2}$ operator at site $i$, and the sum runs over nearest-neighbor pairs $\langle i,j \rangle$. 

In our benchmarks, we use $C_{4v}$ symmetric iPEPS with single unit cell (by flipping the spin on each second site) to approximate the ground state of the model. 
The environment tensors are obtained using the QR-CTMRG method~\cite{zhang2025,yang2025}.
By imposing the $C_{4v}$ symmetry on the local tensors with real numbers, we effectively fix all the local gauge degrees of freedom, thus avoiding the effects of gauge transformations in the optimization~\cite{tang-matrix-2025,tang-gauging-2025-arxiv}.
We use AD to compute the gradient $\partial_A E(A)$.
More specifically, we perform $(3+20)$ steps in the forward computation of QR-CTMRG, in which the last $20$ steps are used for the backpropagation. 
We refer to this as a $(3+20)$ step configuration for contraction method.
For the optimization, we compare the performance of different preconditioning schemes, including the local-metric preconditioner, the full-metric preconditioner and no preconditioning.
The history size used in the L-BFGS algorithm is chosen as $m=200$. The line search employed in all these optimizations is the standard Hager-Zhang line search.
All simulations are initialized from the same random iPEPS tensor and are performed for a maximum of 1000 optimization iterations.
Since we use only a limited number of iterations in QR-CTMRG, the environment tensors and the gradients are accurate only up to a finite precision.  
For this reason, in our benchmarks, we do not aim to push the optimization to full convergence.  
Instead, for each bond dimension, we compare the number of iterations and computational time required to reach a certain reference energy $E_{\mathrm{ref}}$, defined as the lowest energy that can be obtained with the given bond dimension and the number of QR-CTMRG steps used.

\begin{figure}[htbp]
    \centering
    \includegraphics[width=0.48\textwidth]{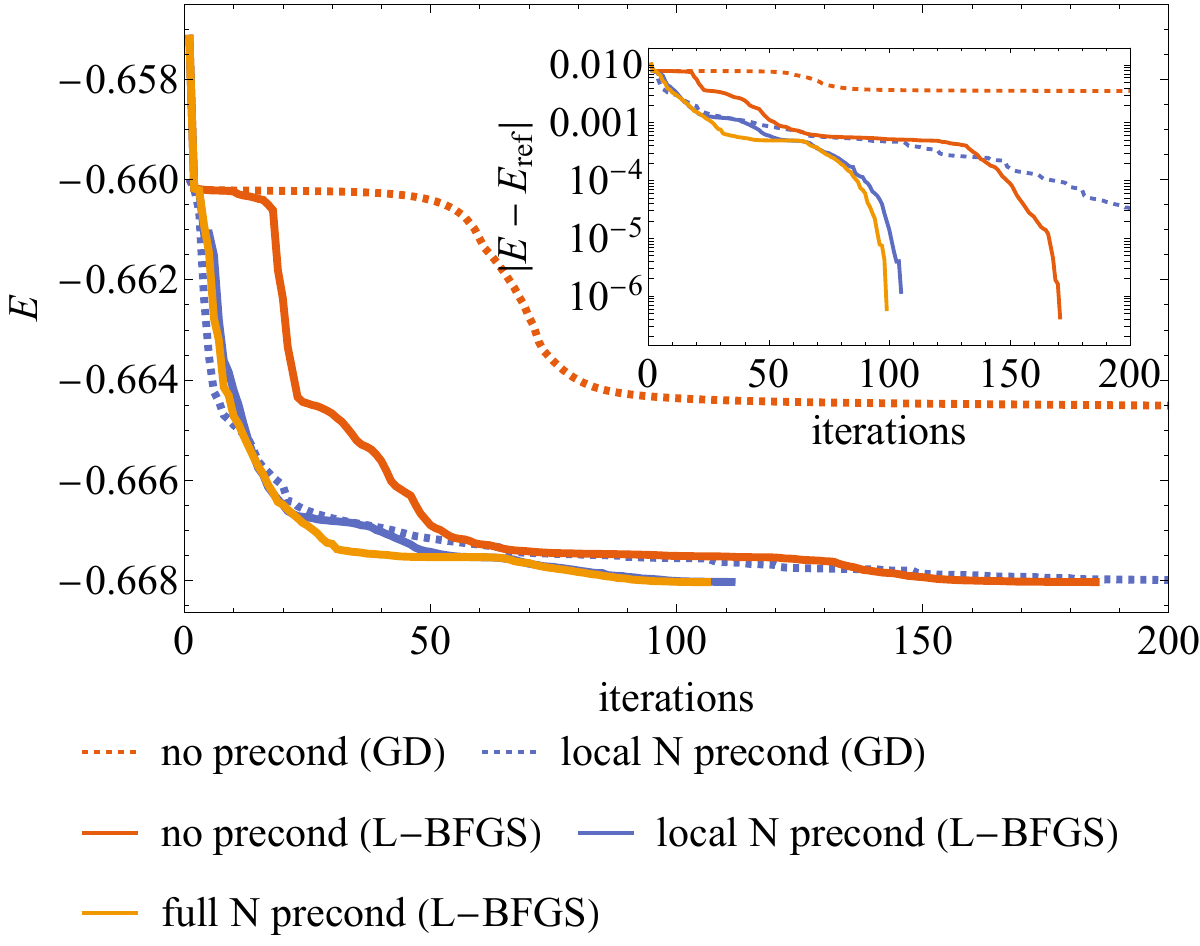}
    \caption{The optimization results for the square-lattice Heisenberg model with $D=3$ iPEPS using different optimization schemes. The inset shows the same data with energy difference to a reference energy on a logarithmic scale.}
    \label{fig:different_methods_compare_iters}
\end{figure}

To start with, we compare the performance of different optimization schemes in $D=3$ iPEPS with environment bond dimension $\chi=50$.
As shown in \cref{fig:different_methods_compare_iters}, we find that, in terms of the number of iteration steps required to reach the same reference energy $E_{\mathrm{ref}}$, using either the local or full metric as a preconditioner can significantly improve the performance of the optimization compared to the non-preconditioned method.  
This effect is particularly pronounced for the simple gradient descent (GD) algorithm, where the preconditioner is truly essential in order for the optimization process to approach the minimum within a reasonable number of iterations. Without preconditioner, the gradient descent algorithm is almost stalling and shows negligible improvement over hundreds or thousands of iterations. This clearly indicates that the employed preconditioner can effectively mitigate the ill-conditioning of the optimization landscape.

The L-BFGS algorithm (without preconditioner) is also able to correct the unfavorable convergence behavior of gradient descent, but still exhibits a number of plateaus where the optimization process is only lowering the energy very slowly. Indeed, the vanilla L-BFGS process is still significantly outperformed by combining L-BFGS with preconditioning.

\begin{figure}[htbp]
    \centering
    \includegraphics[width=0.48\textwidth]{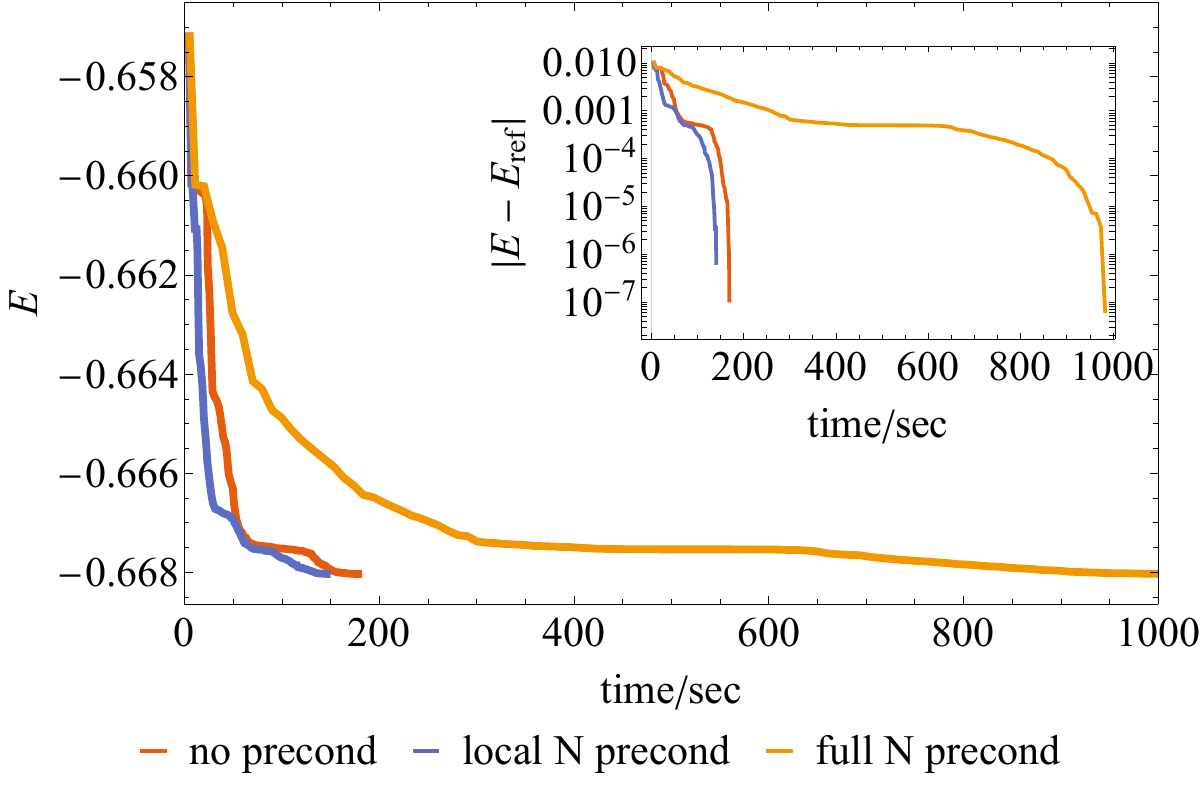}
    \caption{The same optimization results as in \cref{fig:different_methods_compare_iters}, but plotted as a function of wall time. All runtimes are measured on an NVIDIA 4090 GPU.}
    \label{fig:three_methods_compare_time}
\end{figure}

Although \cref{fig:different_methods_compare_iters} shows that the full-metric and local-metric preconditioners achieve similar performance in terms of iteration steps, the computation time per iteration is of course increased, in particular for the full metric.
In \cref{fig:three_methods_compare_time}, we plot the energy along the optimization process as a function of total run time~\footnote{The additional overhead of the L-BFGS algorithm as compared to gradient descent is always negligible in this context, so that using the L-BFGS method (or potentially other quasi-Newton algorithms) is always advantageous, and we do henceforth not include gradient descent results.}. While computing the full-metric preconditioner requires a significant additional cost~\footnote{In the linear problem arising from solving the preconditioned gradient with a full-metric preconditioner, each linear map iteration $(g \mapsto Pg)$ costs roughly the same as a gradient evaluation. This is much more expensive than the corresponding linear map iteration for a local-metric preconditioner, which only involves contractions with local environments.} that annihilates its advantage in reduced number of iterations, the local-metric preconditioner introduces only a small overhead that can largely be neglected, and results in the fastest convergence overall.

\begin{figure}[htbp]
    \centering
    \includegraphics[width=0.48\textwidth]{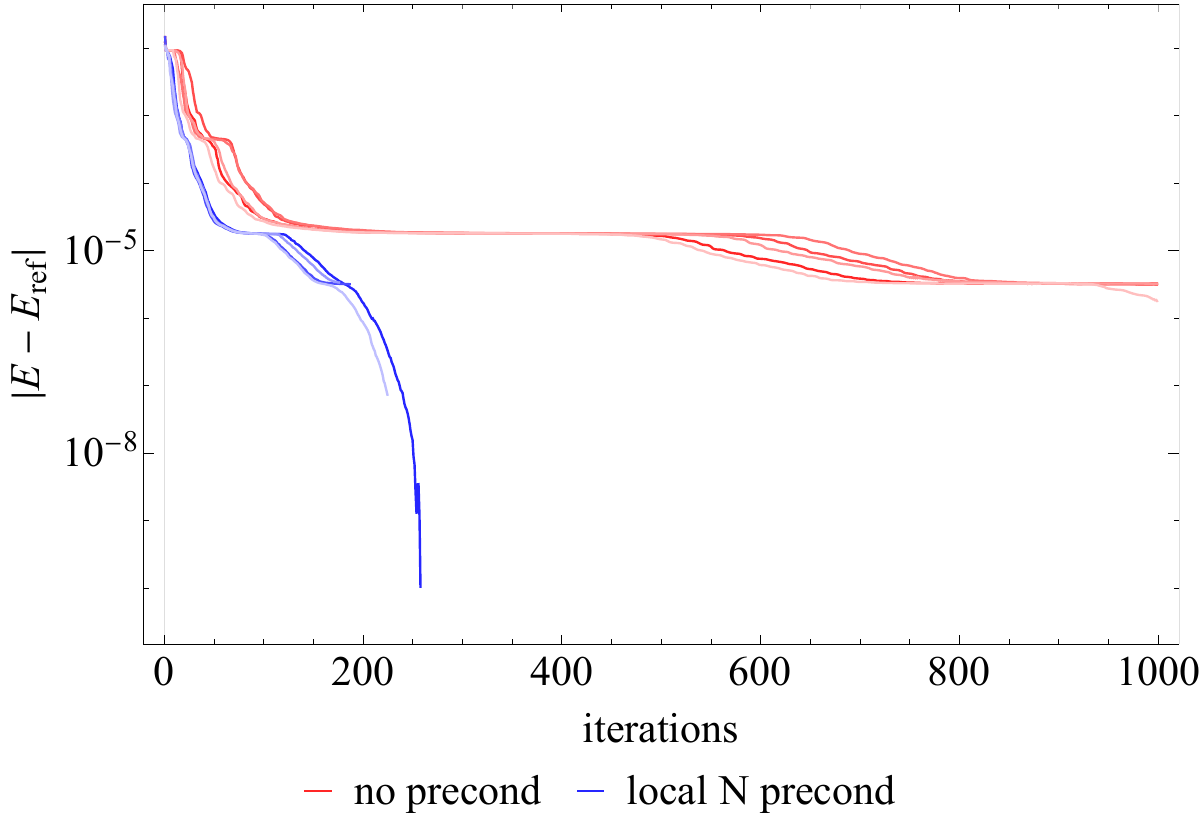}
    \caption{Optimization results for the square-lattice Heisenberg model with $D=5$ using the local-metric preconditioned L-BFGS algorithm and the standard L-BFGS algorithm. 
    We use different shades of lines to indicate different random initializations. 
    }
    \label{fig:NP_D5}
\end{figure}

In fact, the effectiveness of the preconditioning becomes even more pronounced for higher values of the PEPS bond dimension $D$ (and thus the number of variational parameters).
In \cref{tab:heisenberg}, we compare the number of optimization steps and the total computational time required to reach the reference energy $E_{\mathrm{ref}}$ using the local-metric preconditioned L-BFGS algorithm and the standard L-BFGS algorithm. 
Furthermore, in \cref{fig:NP_D5}, we compare the performance of the two optimization schemes for the $D=5$ iPEPS simulation with different initializations of the iPEPS.  
These results demonstrate that the local-metric preconditioned optimization consistently outperforms the non-preconditioned optimization across all tested bond dimensions.  
Overall, the local-metric preconditioner proves to be a practical and effective choice for iPEPS optimization, offering an excellent balance between computational cost and convergence speed.

\begin{table}[htbp]
    \centering
    \begin{tabular}{c|c|c|c|c|c|c}
        \hline
        \hline
        \multirow{2}{*}{$D$} & \multirow{2}{*}{$\chi$} &  \multirow{2}{*}{$E_{\mathrm{ref}}$} & \multicolumn{2}{c|}{steps} & \multicolumn{2}{c}{runtime/s}\\ 
        \cline{4-7} 
          &     &            & no P & P &  no P & P\\
        \hline
        2 &  32 & -0.6602311 & 26      & 11  & 5.17       & 2.83  \\
        3 &  64 & -0.6680561 & 192     & 107 & 236.39     & 168.92\\
        4 & 128 & -0.6692005 & $>1000$ & 194 & $>1561.41$ & 480.60\\
        5 & 160 & -0.6693973 & $>1000$ & 237 & $>2060.00$ & 572.06\\
        6 & 192 & -0.6694226 & $>1000$ & 207 & $>3314.70$ & 846.38\\
        7 & 256 & -0.6694318 & $>1000$ & 304 & $>7252.94$ & 2206.14\\
        \hline
        \hline
    \end{tabular}
    \caption{Optimization results in square-lattice Heisenberg model using iPEPS with larger bond dimensions $D$.
    The table shows the number of optimization steps and computational time to reach $E_{\mathrm{ref}}$ using L-BFGS with or without precondition methods. 
    The reference energy $E_{\mathrm{ref}}$ for each bond dimension is chosen as the lowest energy obtained during the optimization with the corresponding $D$ and $\chi$.
    For $D=7$ case, we show the results using the QR-CTMRG with $(30+30)$ step configuration. All runtimes were measured on an NVIDIA H100 GPU, except for the D=2 case, which was run on an Intel i9-14900KF CPU.
    }
    \label{tab:heisenberg}
\end{table}

Additionally, we compare the performance of the mean-field (BP) preconditioner with that of the local-metric preconditioner, the results of which are presented in Appendix~\ref{sec:BP_preconditioner}.  
Although the BP preconditioner can also significantly accelerate the optimization compared to the non-preconditioned optimization, it appears less reliable and is often outperformed by the local-metric preconditioner.  
Since the additional computational cost of the local metric is already negligible compared to the cost of tensor-network contractions and gradient evaluations, it does not appear beneficial to further reduce the cost of the local metric using BP.

In this regard, we conclude that the local-metric preconditioner appears to be the most effective among those tested, as it achieves a favorable balance between computational cost and preconditioning performance.

\subsection{Results in iPEPS with large unit cell}
A further advantage of the local-metric preconditioner, as compared to the full metric, is that it can be easily extended to iPEPS with large unit cells. 
In this case, we only need to construct and apply the preconditioner for each tensor in the unit cell separately, using its local environment tensors. In contrast, the full metric would couple the different tensors, leading to a coupled linear system for applying the inverse preconditioner to the gradient.

In the following, we benchmark the performance of the preconditioned L-BFGS algorithm on the Heisenberg model on a square lattice with a $2\times2$ unit cell iPEPS and spin-1 Kitaev model on a honeycomb lattice with $2\times6$ unit cell iPEPS in a square lattice~\cite{zhang2025Topological}. 
The Hamiltonian of the Heisenberg model is the same as Eq.~\eqref{eq:H_heisenberg}. 
The Hamiltonian of the spin-1 Kitaev model is given by
\begin{equation}
    H = - \sum_{\langle i,j \rangle_\gamma} S_i^\gamma S_j^\gamma,
\end{equation}
where $S_i^\gamma$ ($\gamma=x,y,z$) are the spin-1 operators at site $i$, and the sum runs over nearest-neighbor pairs $\langle i,j \rangle_\gamma$ connected by $\gamma$-type bonds.

For large unit cell iPEPS, the environment tensors are obtained using the VUMPS method~\cite{nietner2020efficient, PhysRevB.108.085103}. 
When using AD to compute the energy derivative, we perform $(30+4)$ VUMPS iteration steps in the forward computation of VUMPS, using the last 4 steps of which for backpropagation. 
The dominant eigenvector problems in the VUMPS algorithm are solved approximately using a 5-step power method rather than a Krylov method, in order to achieve better performance on GPUs.
We do not impose any gauge fixing scheme on the iPEPS tensors. In order to mitigate the effects of gauge transformations, we make use of a large enough environment bond dimension~\cite{tang-matrix-2025,tang-gauging-2025-arxiv}.
As in Sec.~\ref{sec:localfull}, we do not seek to push the optimization to full convergence, and only compare the number of steps and computational time required for the optimization to reach a certain reference energy. 

The results are shown in \cref{tab:large_unit_cell}. 
For both the Heisenberg model and the spin-1 Kitaev model, the preconditioned L-BFGS algorithm with local-metric preconditioner shows a substantial speed up compared to the standard L-BFGS approach. 
This further confirms the versatility and robustness of the preconditioning approach across different models and lattice geometries.

\begin{table}[htbp]
    \centering
    \footnotesize
    \begin{tabular}{c|c|c|c|c|c|c|c|c}
    \hline
    \hline
    \multirow{2}{*}{model} & \multirow{2}{*}{$D$} & \multirow{2}{*}{$\chi$} & \multirow{2}{*}{unit cell} &  \multirow{2}{*}{$E_{\mathrm{ref}}$} & \multicolumn{2}{c|}{steps} & \multicolumn{2}{c}{runtime/s}\\ 
    \cline{6-9} 
      &     &         &    &  & no P & P &  no P & P\\
    \hline
    Heisenberg & 7 & 256 &$\sqrt{2}\times\sqrt{2}$ & -0.6694302 & 153 & 50 & 64450.30 & 32015.81\\
    Kitaev & 6 & 128 & $2\times6$ & -0.6455729 & 734 & 76 & 21342.36 & 3332.31\\
    \hline
    \hline
    \end{tabular}
    \caption{Optimization results in the square-lattice Heisenberg model and the honeycomb-lattice Kitaev model. 
    The $\sqrt{2}\times \sqrt{2}$ unit cell for the Heisenberg model simulation represents $2\times 2$ unit cell with ABBA pattern. 
    The table shows the number of optimization steps and the computational time to reach $E_{\mathrm{ref}}$ for these models. 
    The reference energy $E_{\mathrm{ref}}$ for each bond dimension is chosen as the lowest energy obtained during the optimization with the corresponding $D$ and $\chi$. All runtimes were measured on two paralleled NVIDIA H100 GPUs. 
    }
    \label{tab:large_unit_cell}
\end{table}

\section{Conclusion and outlook}
In the present work, we have demonstrated that gradient-based optimization of iPEPS can be effectively accelerated by using a preconditioner that is inspired by the tangent-space metric.  
In particular, we show that a local approximation to this metric often strikes the right balance, namely it maintains the good preconditioning quality (thereby significantly reducing the number of iterations) while its application only introduces a negligible computational overhead.
Furthermore, the local preconditioner is easy to implement, because it is composed of the environment tensors that are anyway required to compute expectation values, such as the energy itself, and can be applied to iPEPS with different unit cells and lattice geometries, making it a practical choice for large-scale iPEPS simulations.

We have demonstrated the effectiveness of our approach through numerical benchmarks on different two-dimensional quantum lattice models, including the square-lattice Heisenberg model and the spin-1 honeycomb-lattice Kitaev model. 
In all benchmark examples, the preconditioned optimization shows significant improvements in convergence speed and overall computational efficiency, highlighting the potential of preconditioned optimization methods in advancing tensor network algorithms for strongly correlated systems.

One final observation is that the metric-based preconditioners do not actually take into account any information about the Hamiltonian. This seems at odds with the standard rule that the preconditioner should aim to provide a low-cost (and easily invertible) approximation to the Hessian of the cost function ---the variational energy--- and raises the question whether alternative preconditioners can be conceived that do explicitly take contributions from the Hamiltonian into account. Whether, in the context of tensor networks and iPEPS in particular, such a preconditioner can be constructed and applied with a computational cost that does not defeat its very purpose remains an open problem. Indeed, even in the simpler case of matrix product states, no such strategy has been developed.

\section{Acknowledgements}
We thank Lei Wang, Yuchi He, and Laurens Vanderstraeten for helpful discussions. 
X.Y.Z. and J.H. are supported by European Research Council (ERC) under the European Union's Horizon 2020 program (grant agreement No. 101125822). 
W.T. is supported by Research Foundation Flanders (FWO) [Postdoctoral Fellowship 12AA225N]. 
Q.Y. and P.C. are supported by the European Research Council (ERC) under the European Union's Horizon 2020 programme (grant agreement No. 101001604).
For code developments and numerical calculations, we thank EuroHPC Joint Undertaking for awarding us access to MareNostrum5 at BSC, Spain under Grant No. EHPC DEV-2025D07-003 and DEV-2025D06-052 and access to LUMI, Finland under Grant No. VSC-2025-09-T350-JH-JULIA.

\paragraph*{Data availability.}
To streamline practical implementation, we provide a Julia implementation~\cite{iPEPS_preconditioner} for reference, along with benchmark data~\cite{data} for comparison.
Both the standard and preconditioned gradient-descent and L-BFGS algorithms used in the iPEPS optimization are implemented in the open-source package \texttt{OptimKit.jl}~\cite{OptimKit.jl}.

\appendix

\section{BP preconditioner}
\label{sec:BP_preconditioner}
\begin{figure}[!htbp]
    \centering
    \includegraphics[width=0.4\textwidth]{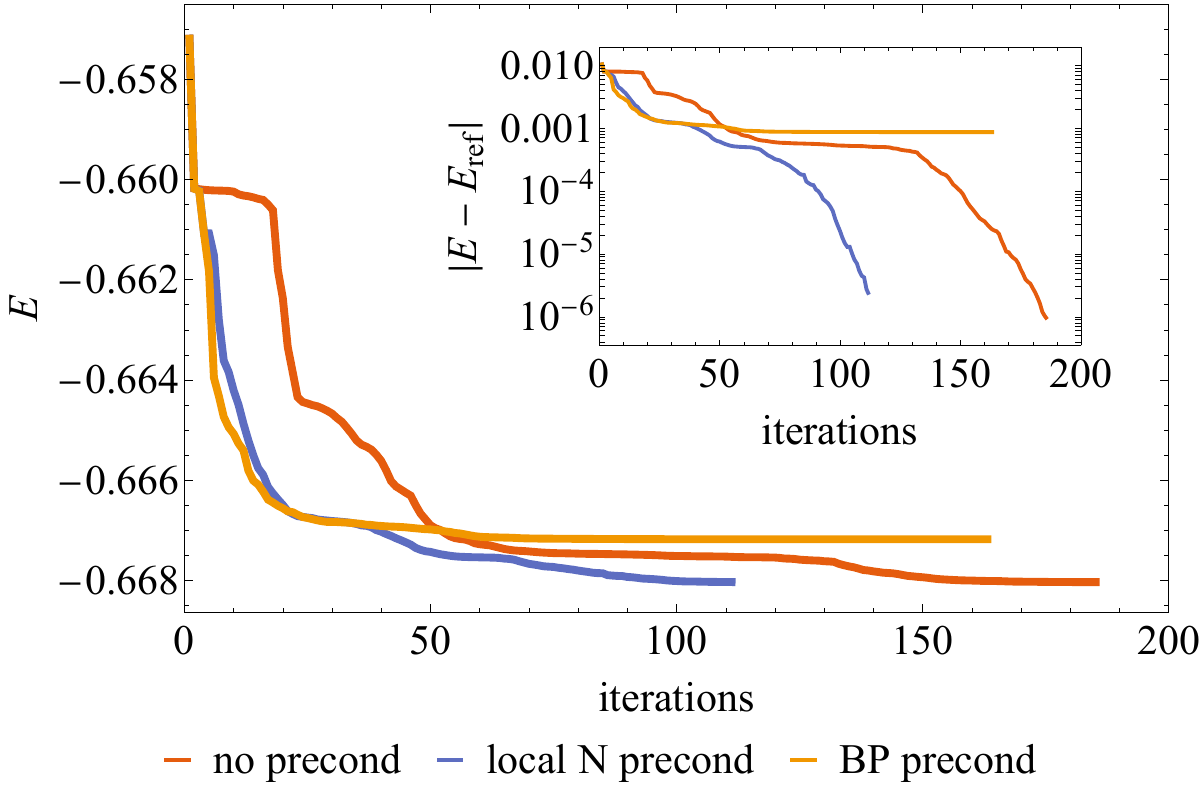}
    \includegraphics[width=0.4\textwidth]{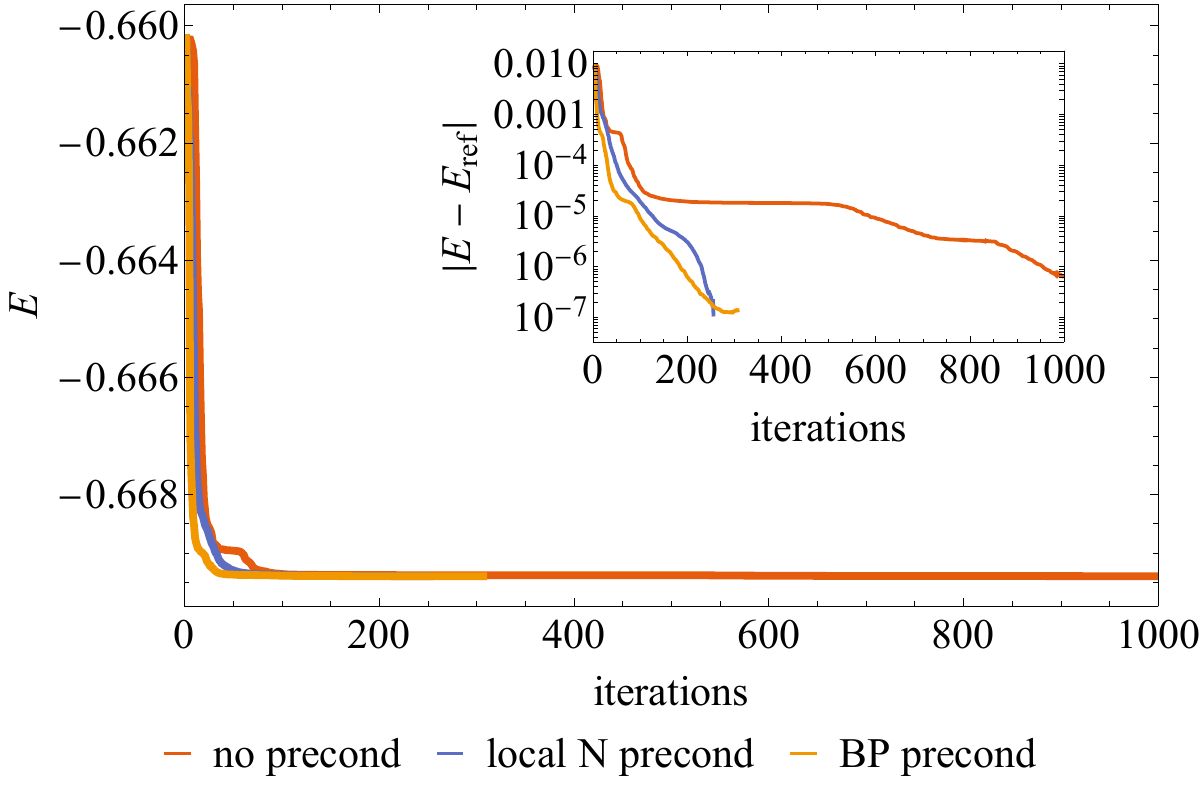}
    \includegraphics[width=0.4\textwidth]{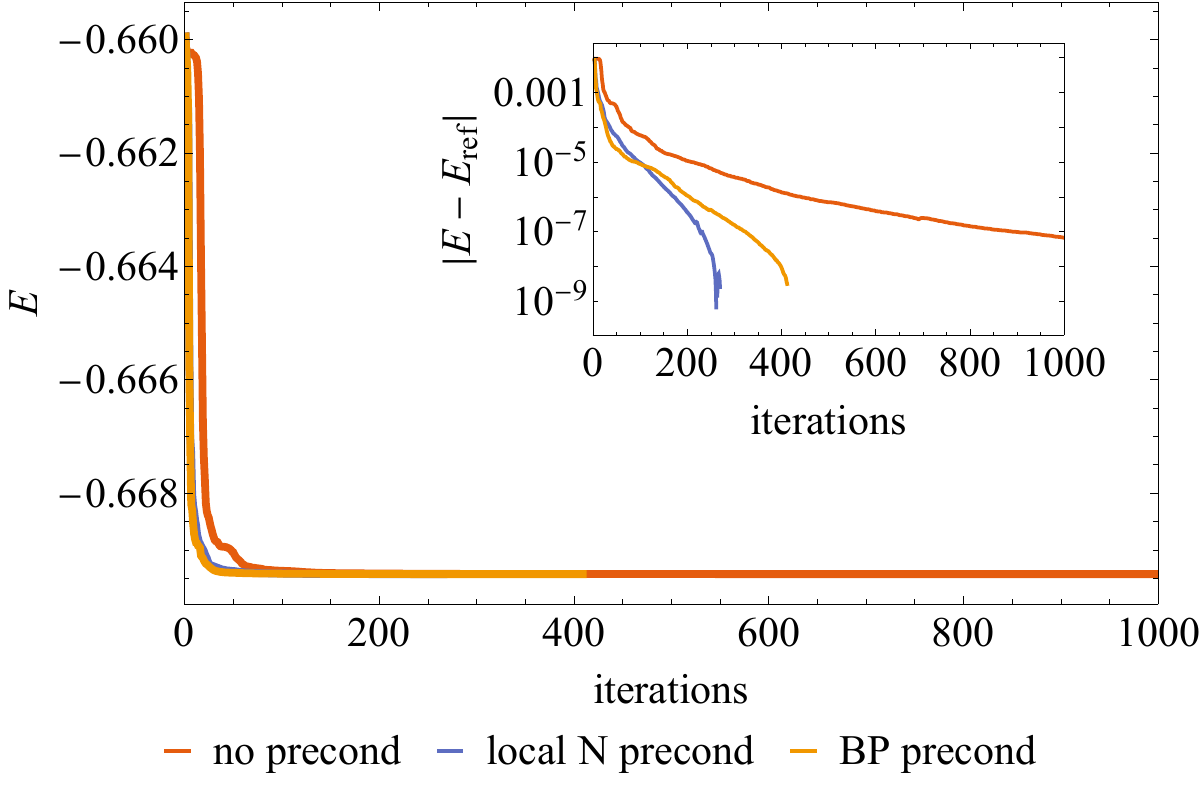}
    \caption{Optimization results for the square-lattice Heisenberg model using different preconditioners.
    From top to bottom, the bond dimensions of the iPEPS are $D=3, 5,$ and $7$, respectively.
    The inset shows the same data with energy difference to a reference energy on a logarithmic scale.}
    \label{fig: BP_metric}
\end{figure}

Using the same setup as in \cref{sec:localfull}, we compare the BP preconditioner against the local metric preconditioner and a standard L-BFGS method without preconditioning. 
As shown in \cref{fig: BP_metric}, the BP preconditioner can also significantly accelerate the optimization compared to the non-preconditioned optimization. 
However, due to the drastic approximations made in the metric computation, we find that its performance is not consistent. While it performs on par with the local-metric preconditioner in some cases, we also find cases where it causes the optimization to stall, or where it is just clearly outperformed by the local-metric preconditioner, especially at larger bond dimension.

\bibliography{refs.bib}
\end{document}